\documentclass[nonacm,sigconf]{acmart}
\AtBeginDocument{%
  }

\setcopyright{acmlicensed}
\copyrightyear{2018}
\acmYear{2018}
\acmDOI{XXXXXXX.XXXXXXX}

\acmConference[Conference acronym 'XX]{Make sure to enter the correct
  conference title from your rights confirmation emai}{June 03--05,
  2018}{Woodstock, NY}
\acmISBN{978-1-4503-XXXX-X/18/06}

\usepackage{multirow} 

\begin{document}

\title{From Critique to Clarity: A Pathway to Faithful and Personalized Code Explanations with Large Language Models}

\author{Luo Zhang}\authornotemark[1]
\affiliation{%
  \institution{Worcester Polytechnic Institute}
    \state{MA}
  \country{USA}}
\email{zluo3@wpi.edu}
\author{Zexing Xu}\authornotemark[1]
\affiliation{%
  \institution{University of Illinois Urbana-Champaign}
      \state{IL}
  \country{USA}}
\email{zexingx2@illinois.edu}
\author{Yichuan Li}
\authornote{Authors contributed equally to this research.}
\affiliation{%
  \institution{Worcester Polytechnic Institute}
      \state{MA}
  \country{USA}}
\email{yli29@wpi.edu}

\author{Seyed Rasoul Etesami}
\affiliation{%
  \institution{University of Illinois Urbana-Champaign}
        \state{IL}
  \country{USA}}
\email{etesami1@illinois.edu}

\author{Kyumin Lee}
\affiliation{%
  \institution{Worcester Polytechnic Institute}
        \state{MA}
  \country{USA}}
\email{kmlee@wpi.edu}


\begin{abstract}
In the realm of software development, providing accurate and personalized code explanations is crucial for both technical professionals and business stakeholders. Technical professionals benefit from enhanced understanding and improved problem-solving skills, while business stakeholders gain insights into project alignments and transparency. Despite the potential, generating such explanations is often time-consuming and challenging. This paper presents an innovative approach that leverages the advanced capabilities of large language models (LLMs) to generate faithful and personalized code explanations. Our methodology integrates prompt enhancement, self-correction mechanisms, personalized content customization, and interaction with external tools, facilitated by collaboration among multiple LLM agents. We evaluate our approach using both automatic and human assessments, demonstrating that our method not only produces accurate explanations but also tailors them to individual user preferences. Our findings suggest that this approach significantly improves the quality and relevance of code explanations, offering a valuable tool for developers and stakeholders alike.
\end{abstract}

\ccsdesc[500]{Computing methodologies~Natural language generation}

\keywords{Code Explanation, Large Language Models, Personalization, Prompt Engineering}


\maketitle

\section{Introduction}
Code explanations are crucial in the digital landscape, serving as essential learning tools for tech professionals and aligning technical projects with business goals for stakeholders \citep{feng2020codebert, husain2019codesearchnet, guo2022unixcoder}.
Efforts to address these code understanding challenges have included tasks such as code summarization and code comment generation. Code summarization provides high-level overviews but often lacks detailed insights, making it primarily useful for documentation purposes \citep{leclair2019neural, allamanis2018survey}. Conversely, code comment generation involves line-by-line commenting, offering detailed explanations but potentially overwhelming users seeking a broader understanding \citep{hu2018deep, li2018automatic}. 
The diversity of user needs—ranging from data scientists requiring domain-specific insights to software engineers focusing on architectural details—necessitates a more tailored approach to personalized code explanation. 
Personalized code explanations tailored to users' backgrounds and knowledge levels are essential for effectively conveying complex information \citep{alhuzali2021model, dan2021improving}. However, creating comprehensive, in-depth, and personalized explanations is time-consuming and resource-intensive \citep{miceli2021studying, linnell2019generating}, limiting their availability and posing challenges for both learners and developers.
Besides, Faithfulness which is also very important in code explanation. It ensures that the generated text accurately reflects the code's functionality and logic, avoiding any misinterpretation or oversimplification \citep{atanasova2020diagnostic, hase2021evaluating}. 
\begin{figure*}[tbh!]
  \includegraphics[width=\textwidth]{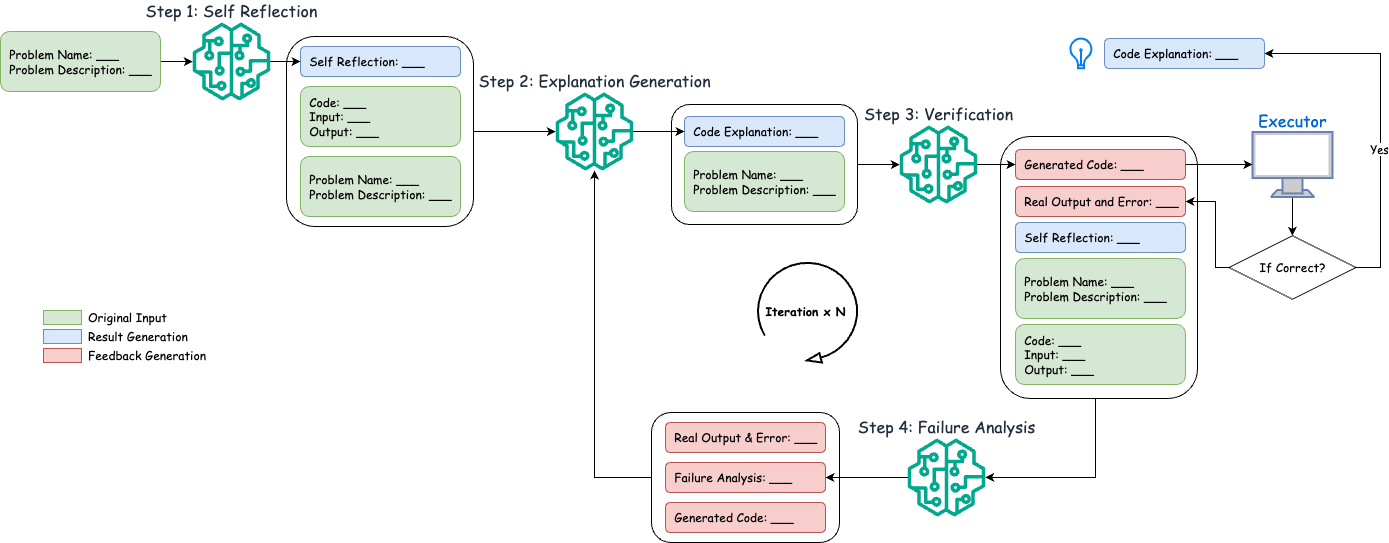}
      \caption{The Illustration of Iterative Code Explanation Refinement. The system generates code explanations through two iterative loops: a faithfulness loop to ensure technical accuracy, and a personalization loop to tailor the explanation to the user's background. The loops operate independently to optimize their respective objectives and navigate the potential trade-off between personalization and faithfulness}
  \Description{The system generates code explanations through two iterative loops: a faithfulness loop to ensure technical accuracy and a personalization loop to tailor the explanation to the user's background. The loops operate independently to optimize their respective objectives and navigate the potential trade-off between personalization and faithfulness.}
  \label{fig:teaser}
\end{figure*}
Recent advancements in large language models (LLMs) offer a promising solution to the challenges of code explanation. LLMs have demonstrated exceptional performance across a diverse array of tasks, primarily due to their enhanced reasoning capabilities \citep{brown2020language, lewkowycz2022solving}. The efficacy of LLMs in tackling complex tasks heavily relies on advanced prompt engineering techniques, such as chain-of-thought (CoT) prompting \citep{wei2022chain}. This method, along with iterative prompting and question decomposition, enhances the logical flow and clarity of explanations \citep{paul2023refiner, weng2022large, madaan2024self, hou2023promptboosting, pitis2023boosted, gou2023critic}. These approaches collectively contribute to the substantial promise in improving the accuracy and effectiveness of LLM-generated responses in solving complex problems. However, generating personalized and faithful code explanations requires more than a single prompt~\citep{Bhattacharya2023ExploringLL, osti_10397860}; it necessitates complex and sequential prompt design.

Current methods for code explanation face significant challenges in providing faithful, personalized explanations, and balancing various requirements. Many approaches struggle with \textbf{faithfulness}, often producing explanations that are syntactically correct but semantically incorrect, leading to misunderstandings or errors in code interpretation \citep{lin2022program}. For example, an LLM might describe a sorting algorithm correctly in terms of its steps but fail to explain its actual complexity or edge cases. Ensuring \textbf{personalization} is another challenge, as generic explanations that ignore the user's background, expertise, or specific needs result in less effective communication \citep{ahmad2022contextual}. For instance, novice programmers might require step-by-step explanations, while experienced developers might prefer summaries. Additionally, balancing accuracy, completeness, and personalization remains difficult ~\citep{Bhattacharya2023ExploringLL, osti_10397860}. Detailed explanations might overwhelm users, while brief ones might omit crucial information, making it challenging to strike the right balance for effective code explanations. \citep{milivcka2024large} also found that LLMs can downplay their cognitive abilities to fit the personas they simulate.

To this end, we propose an innovative iterative refinement approach that integrates prompt augmentation, self-correction mechanisms, and personalized content adaptation based on user preferences. 
By incorporating prompt augmentation \citep{reynolds2021prompt, liu2023pre}, our approach enriches initial prompts with additional context and hints, guiding the LLM towards generating more accurate, relevant, and detailed explanations.
Our method includes self-correction mechanisms \citep{madaan2023selfrefine, press2022self}, which iteratively improve response quality through feedback and correction processes. This ensures stable and accurate outputs by continuously refining content through multiple iterations, leading to highly faithful explanations.
Additionally, by leveraging personalized content adaptation based on user preferences \citep{zamfirescu2023iterative, madotto2019personalizing}, our approach delivers more engaging and tailored explanations. Analyzing users' historical interactions and preferences allows us to align explanations with their unique needs, such as their tendency to ask detailed questions or their preference for high-level summaries. Incorporating examples and application studies relevant to the user's domain and interests further enhances understanding and engagement. 
 
Our framework also leverages the strengths of LLMs by incorporating external tools and collaboration among diverse LLM agents \citep{chen2024comm}. This integration enhances the comprehensiveness and accuracy of explanations, addressing the multifaceted requirements for high-quality code explanations.

To evaluate the effectiveness of our proposed method, we conducted extensive experiments on the CodeContests dataset \citep{li2022competition} from Codeforces\footnote{\url{codeforces.com}}. We employed a combination of automatic and human evaluation metrics to assess the quality of the generated explanations
. Our experimental results consistently demonstrate that our method produces more accurate and personalized code explanations than existing approaches. 
The main contributions of this work are threefold:
\begin{itemize}
\item We introduce the novel task of personalized code explanation generation using LLMs, addressing the challenge of balancing faithfulness and adaptability to individual user preferences.
\item We propose an innovative iterative refinement approach that integrates prompt augmentation, self-correction, and personalized content adaptation, leveraging LLMs, external tools, and multi-agent collaboration.
\item Our method achieves state-of-the-art performance on the Code-Contest dataset, consistently outperforming existing approaches in generating accurate and personalized code explanations across automatic and human evaluations.
\end{itemize}

\section{Problem Definition}
Our research is based on a formalized code problem dataset, consisting of $n$ individual problems. Each problem, denoted as $p$, is linked to a single human-generated oracle solution, $s$. To generate a faithful explanation $e$ for a problem-solution pair $(p, s)$, we sample from the model's distribution $\mathbb{P}_{\mathcal{M}}$ conditioned on the prompt $\wp$, problem $p$, and solution $s$:

\begin{equation}
e \sim \mathbb{P}_{\mathcal{M}}(\cdot | \wp \oplus p \oplus s)
\end{equation}

Subsequently, given the problem-solution pair $(p, s)$, the faithful explanation $e$, and a user's historical \textit{Stack Overflow} inquiries $h$, we personalize the explanation as $pe$:

\begin{equation}
pe \sim \mathbb{P}_{\mathcal{M}}(\cdot | \wp \oplus p \oplus s \oplus e \oplus h)
\end{equation}

The final output, $o$, combines both $e$ and $pe$:

\begin{equation}
o = e \oplus pe
\end{equation}

\section{Method}
Figure \ref{fig:teaser} provides an overview of the method. Inspired by the iterative approach humans adopt in writing, our approach introduces an iterative explanation refinement process, which is adapted into two specific loops: the faithfulness loop and the personalization loop. These loops work in tandem to generate a code explanation that is both technically faithful and personalized to the user's background and programming skills. It is worth noting that we observed a potential trade-off between pursuing personalization and maintaining faithfulness simultaneously, as optimizing both may lead to compromises. \citet{milivcka2024large} also found that LLMs can downplay their cognitive abilities to fit the personas they simulate. Therefore, we designed these loops as independent components to maximize each objective.

\subsection{Iterative Explanation Refinement}
Drawing inspiration from the iterative refinement employed by humans in writing, our proposed methodology for generating high-quality code explanations consists of a three-stage process: reflection, iterative explanation, and verification and analysis. This systematic approach ensures continuous improvement by detecting and rectifying errors that arise from real-world interactions. 
Although both the faithfulness and personalization loops leverage these shared stages, they adapt them to meet their distinct objectives, reflecting the tailored nature of each loop.

\paragraph{Reflection}
In the reflecting stage, the method leverages the summarization capabilities of LLMs \citet{jin2024comprehensive} to efficiently extract key information from a given context, such as a problem description or a user's historical inquiries on \textit{Stack Overflow}. The output from the reflection stage serves as input for more complex tasks, a crucial component of the subsequent stages' requirements. Thus, the reflection stage plays a crucial role in knowledge accumulation and progression, providing the necessary insights and information to support subsequent, more challenging steps \citet{ridnik2024code}.

\paragraph{Initialization and Refinement} 
This stage, pivotal for enhancing code explanations, entails two key tasks: initial setup and iterative refinement. To commence, we provide the LLM with the code solution, context, and knowledge from the Reflection stage as input. And utilize the chain-of-thought (CoT) methodology described by \citet{wei2022chain} to initiate code explanations. For example, in the case of the faithful loop, the LLM first generates a detailed, sequential explanation and then provides a high-level understanding of the code. If the initial explanation fails to meet certain criteria, a revision process ensues. Here, the LLM is fed with the code, context, the previously generated explanation, and knowledge from the \textit{Verification and Analysis} stage, which is instrumental in pinpointing errors and offering actionable suggestions through external tool interactions. This iterative process continues until predetermined stopping conditions are met, ensuring continuous improvement in the explanation quality.

\paragraph{Verification and Analysis}
Recent studies have demonstrated the capacity of LLMs to interact with external tools \citet{yuan2024easytool, paranjape2023art, wu2023autogen}, enhancing their ability to scrutinize and refine their initial responses. The central concept of this stage involves the LLM engaging with external utilities, such as a Python executor or another LLM, to verify the previously generated explanation. If the external tool output indicates that the previous generated explanation satisfies specific criteria, the refining loop terminate. Otherwise, the LLM is required to analyze the error and provide some revision suggestions for the following explanation improvement.

\subsection{Faithfulness Loop}
The 3-stage refinement is suitable for this process, but some adjustments are necessary. In the \textit{Reflection} stage, given the complex and intricate code problem $p$, we ask the LLM to extract the problem goals, inputs, outputs, conditions and other relevant details, represented as $pr$. 

In the \textit{Iterative Explanation} stage, the initialization step generates an initial step-by-step description and high-level explanation $e_0$ based on the problem $p$, the accepted code solution $s$, and the problem reflection $pr$ generated by the reflecting stage. The revision step generates an improved explanation $e_{i+1}$ based on the problem $p$, the code solution $s$, the problem reflection $pr$, the previous explanation $e_i$, the verification code solution $vs_i$, the executor output $eo_i$, and the failure analysis $a_i$.

In the \textit{Verification and Analysis} stage, to verify if the code explanation is faithful, we test how much it can aid in solving the problem by utilizing the code generating ability of LLMs \citet{li2022competition, ridnik2024code, ni2023lever}. Given the problem $p$ and the previous explanation $e_i$, a verification code solution $vs_i$ is generated. Then verification code solution is executed to obtain the output $eo_i$, which is compared against the public test cases. If the output is incorrect, the LLM analyzes the error and generates an analysis $a_i$ based on the problem $p$, the code solution $s$, the problem reflection $pr$, the verification solution code $vs_i$, and the executor output $eo_i$. We found that LLMs excel more in finding code-related issues compared to textual problems. Therefore, during error analysis, we task the LLM to analyze errors in the verification solution code, and based on this analysis, we modify the code explanation.

\subsection{Personalization Loop}
In addition to faithfulness, the acceptance of code explanations by the audience is crucial. People with different backgrounds and programming skills have varying requirements for code explanations. Therefore, we need this step to produce personalized code explanations.
Different from existing studies \citep{chen2024persona, wang2023rolellm, shao2023characterllm, li2023chatharuhi} on role-playing LLMs, which focus on using demographic tags and conversation history data to simulate personas, our approach leverages users' actions—specifically their inquiry history about Python, data structures, and algorithms on \textit{Stack Overflow}—to infer their programming profile. This method allows the LLM to represent their personas and generate personalized code explanations that align well with their profiles.

The 3-stage refinement works well for this process, although some modifications are required. In the \textit{Reflection} stage, the LLM extracts the user's programming profile $up$ based on their historical inquiries $h$. The user's profile is divided into six aspects: programming languages, skill level, topics of interest, problem-solving approach, experience and other relevant information.

In the \textit{Iterative Explanation} stage, the initialization step generates an initial personalized explanation $pe_0$ based on the problem $p$, the code solution $s$, the user's programming profile $up$, and the explanation $e$ from the faithfulness loop. The revision step generates an improved personalized explanation $pe_{i+1}$ based on the problem $p$, the code solution $s$, the explanation $e$ from the faithfulness loop, the user's profile $up$, the previous personalized explanation $pe_i$, and the rating $r_i$ on the previous personalized explanation that is generated by a role-playing LLM.

In the \textit{Verification and Analysis} stage, the role-playing judging LLM evaluates whether the previous personalized explanation $pe_i$ aligns well with the user's programming skills and background, based on the user's profile $up$, the problem $p$, the code solution $s$, and the personalized explanation $pe_i$. The LLM generates a rating $r_i$. If $r_i$ indicates that the role-playing LLM is not satisfied, the LLM also provide some revision suggestions. Otherwise, the loop terminates.

After the faithfulness and personalization loops, we combine the $e$ and the $pe$ together as our final output $o$.

\section{Experiment Setup}
\paragraph{Code Problem Dataset} This study utilizes the CodeContests dataset \cite{li2022competition}, sourced from competitive programming platforms such as Codeforces, to ensure the robustness and validity of our findings. To mitigate data leakage, we exclusively rely on the validation and test sets as our primary data sources. We rigorously filter out problems with image-based descriptions and those lacking oracle Python solutions. The validation set comprises 67 authentic contest problems collected from various online platforms, while the test set consists of 102 instances. Moreover, each problem in the dataset includes multiple oracle solutions. 
Given the constraints on the context window size of LLMs \cite{zhao2023survey}, we adopt the shortest solution for each problem.

\paragraph{Inquiry History Dataset}
To collect the real user coding preference, we collected user profiles from \textit{Stack Overflow}\footnote{\url{stackoverflow.com/}} using an anonymized dump of all user-contributed content on the website \cite{rae2021scaling}, which includes questions, answers, comments, tags, and related data. 
Specifically, we choose 10 users from the dataset and for each user, we sampled their five most recent inquiries related to Python, data structures, and algorithms, which include the title, tags, and body of each inquiry.

\paragraph{Baseline}
Several studies have investigated the application of LLMs in generating code explanations \citet{li2023explaining, macneil2023experiences, brusilovsky2023explaining, sarsa2022automatic, chen2023gptutor, leinonen2023comparing, macneil2022generating, oli2023behavior}, yet their approaches often center on the feasibility of LLMs with simple prompts. \citet{li2023explaining}'s work stands out for its emphasis on high-quality code explanation generation, which we adopt as our primary reference. 
The baseline method relies on a naive greedy decoding strategy, which we enhance by integrating self-consistency principles from \citet{wang2023selfconsistency}, known to improve response quality. We also propose a strong baseline, \textit{Self-Selection}, where the LLM generates $n$ explanations and then ranks them based on predefined criteria, simulating a decision-making process to select the most suitable explanation.
Both the baseline and \textit{Self-Selection} aim to produce explanations that are both factually accurate and personalized for a specific code solution.

\paragraph{Backbone Model} 
During the generation phase, we employed GPT-3.5-turbo \citet{openaigpt3.5} in both the faithfulness and personalization loops. For the evaluation, GPT-3.5-turbo was utilized exclusively. In code generation, a temperature of 0.2 and a top-p probability of 0.1 were set, while for text generation, the temperature was set to 0.7 and the top-p was set to 0.8.

\paragraph{Other Setting}
Recent studies have demonstrated that after undergoing 3 to 4 \citet{gou2023critic}, LLMs are capable of producing higher-quality responses. Therefore, we established the iteration count as 4.  For fairly comparing, in the \textit{Self-Selection} method, we also ask the LLM to sample 4 responses for the same instruction.

\subsection{Automatic Evaluation Metric}

\paragraph{Pass@k} 
This metric evaluates how well explanations generated by the LLM solve code problems \cite{chen2021evaluating, li2023explaining, ridnik2024code}. We assess if the LLM can solve a problem using its generated explanations by sampling $k$ programs and measuring their solve rate@k against ground-truth outputs, derived from private or generated test cases..

\paragraph{Win Rate} 
This metric evaluates personalized explanations generated by different methods using the win rate metric \cite{wang2023rolellm}. It compares how often one explanation outperforms another when simulating an individual with a specific user profile.

\paragraph{Rouge-L}
This metric evaluates personalized explanations by measuring overlap between model predictions and user queries on Stack Overflow \cite{lin2023generating, lin2004rouge}. A higher value indicates better alignment with the user's skill level and background.

\paragraph{Word Overlap Ratio}
This metric evaluates personalized explanations by measuring word overlap between generated content and user queries on Stack Overflow, indicating similarity with the user's profile.
\\

To avoid generation uncertainty of LLMs \citet{lin2023generating}, we sample 4 times for each problem and each chosen method. Thus, for each metric above, we report the average value over 40 calculations (10 chosen users * 4 samples per user). 

\begin{table*}[ht]
    \centering
        \begin{tabular}{c|c|c|cc} 
        \hline
            Model & Set & Method & Pass@1 & Pass@5 \\
            \hline
            \multirow{8}{*}{GPT-3.5} & \multirow{3}{*}{Validation} & Baseline & 25.11\% \footnotesize{$\pm$ 0.0369} & 29.29\% \footnotesize{$\pm$ 0.0347} \\
             &  & Self-Selection (Sample n = 4) & 26.08\% \footnotesize{$\pm$ 0.0312} & 30.34\% \footnotesize{$\pm$ 0.0327} \\
             &  & Self-Iteration (Iteration n = 4) & \textbf{30.11\% \footnotesize{$\pm$ 0.0313}} & \textbf{34.89\% \footnotesize{$\pm$ 0.0298}} \\
            \cline{2-5} 
             & \multirow{5}{*}{Test} & Baseline & 21.57\% \footnotesize{$\pm$ 0.0284} & 25.49\% \footnotesize{$\pm$ 0.0253} \\
             &  & Self-Selection (Sample n = 4) & 22.60\% \footnotesize{$\pm$ 0.0275} & 26.45\% \footnotesize{$\pm$ 0.0264} \\
             &  & Self-Iteration (Iteration n = 4) & \textbf{26.52\% \footnotesize{$\pm$ 0.0291}} & \textbf{30.15\% \footnotesize{$\pm$ 0.0262}} \\
             &  & Commercial Product 1 & 16.67\% & 17.65\% \\
             &  & Commercial Product 2 & 29.41\%  & 29.41\% \\
        \hline
        \end{tabular}
    \caption{Pass@k}
    \label{tab:Pass@k}
\end{table*}

\begin{figure*}[htbp]
    \centering
    {\includegraphics[width=\linewidth]{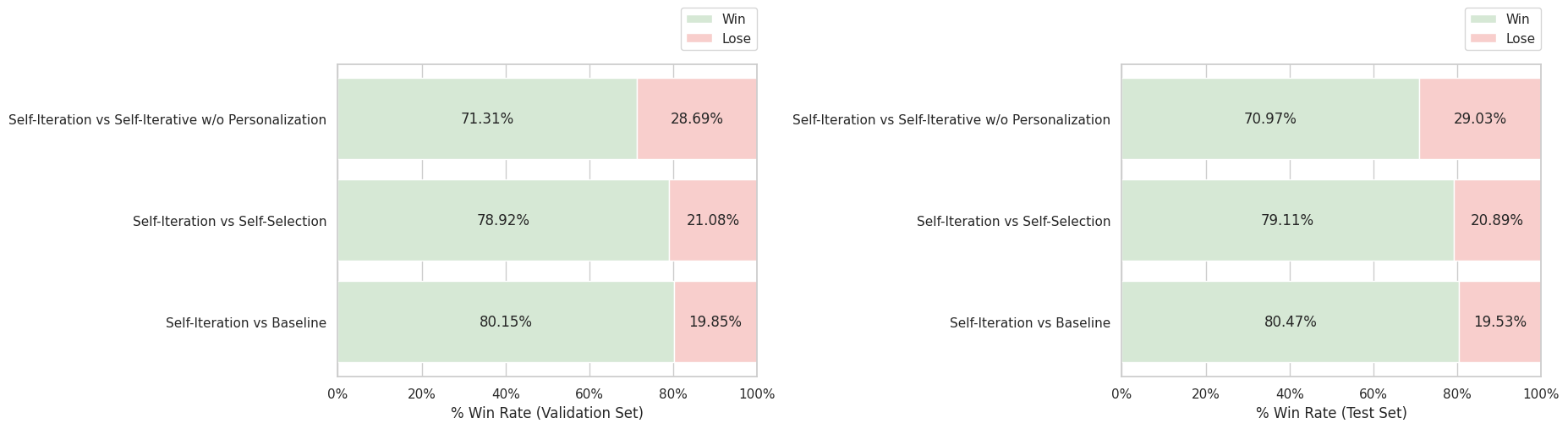}}
    \caption{Win Rate}
    \label{fig:win_rate}
\end{figure*}

\begin{table}[ht]
    \centering
    \resizebox{\linewidth}{!} {
    \renewcommand{\arraystretch}{1.2}
        \begin{tabular}{c|c|c|c}
        \hline
            Model & Set & Method & Rouge-L \\
            \hline
            \multirow{6}{*}{GPT-3.5} & \multirow{3}{*}{Validation} & Baseline &  0.0230 \footnotesize{$\pm$ 0.0085} \\
             &  & Self-Selection & 0.0232 \footnotesize{$\pm$ 0.0086} \\
             &  & Self-Iteration & \textbf{0.0363 \footnotesize{$\pm$ 0.0133}} \\
            \cline{2-4} 
             & \multirow{3}{*}{Test} & Baseline & 0.0227 \footnotesize{$\pm$ 0.0083} \\
             &  & Self-Selection & 0.0230 \footnotesize{$\pm$ 0.0085} \\
             &  & Self-Iteration & \textbf{0.0361 \footnotesize{$\pm$ 0.0131}} \\
        \hline
        \end{tabular}
    }
    \caption{Rouge-L}
    \label{tab:rouge_l}
\end{table}

\begin{table}[ht]
    \centering
    \resizebox{\linewidth}{!} {
    \renewcommand{\arraystretch}{1.2}
        \begin{tabular}{c|c|c|c}
        \hline
            Model & Set & Method & Word Overlap Ratio \\
            \hline
            \multirow{6}{*}{GPT-3.5} & \multirow{3}{*}{Validation} & Baseline & 4.99\% \footnotesize{$\pm$ 0.0113} \\
             &  & Self-Selection & 5.04\% \footnotesize{$\pm$ 0.0114} \\
             &  & Self-Iteration & \textbf{7.44\% \footnotesize{$\pm$ 0.0150}} \\
            \cline{2-4} 
             & \multirow{3}{*}{Test} & Baseline & 4.86\% \footnotesize{$\pm$ 0.0112} \\
             &  & Self-Selection & 4.90\% \footnotesize{$\pm$ 0.0110} \\
             &  & Self-Iteration & \textbf{7.35\% \footnotesize{$\pm$ 0.0148}} \\
        \hline
        \end{tabular}
    }
    \caption{Word Overlap Ratio}
    \label{tab:word_overlap_ratio}
\end{table}

\section{Results and Analysis}
In this section, we present and analyze the results from two key angles: faithfulness and personalization. These angles provide insight into how well the generated explanations assist in solving code problems and how effectively they cater to individual users' profiles.

\paragraph{Faithfulness}
Faithfulness is assessed through the Pass@k metric, which measures the success rate of the generated explanations in solving coding problems.

The results of Pass@k, summarized in Table \ref{tab:Pass@k}, highlights the superior performance of the \textit{Self-Iteration} method, which consistently surpasses both the baseline and \textit{Self-Selection} techniques in both validation and test sets. Notably, our approach even outperforms several specialized online commercial products specifically tailored for code explanation. Given the absence of personalized code explanation features in these online products, our assessment centers on their faithfulness in code interpretation. This indicates that the iterative refinement process of the \textit{Self-Iteration} method significantly enhances the accuracy and utility of the explanations. The method's ability to iteratively improve responses leads to explanations that are more reliable and effective in solving code problems.

\paragraph{Personalization}
Personalization is evaluated through the Win Rate, Rouge-L, and Word Overlap Ratio metrics, focusing on how well the generated explanations align with individual user profiles.

The Win Rate results, illustrated in Figure \ref{fig:win_rate}, show that the \textit{Self-Iteration} method substantially outperforms other methods. This highlights the effectiveness of iterative refinement in tailoring explanations to individual users' preferences and programming skills. The \textit{Self-Iteration} method's ability to adapt explanations based on user profiles leads to more personalized and contextually appropriate content.

Table \ref{tab:rouge_l} and Table \ref{tab:word_overlap_ratio} summarize the results for Rouge-L and Word Overlap Ratio metrics. The \textit{Self-Iteration} method consistently achieves higher scores, indicating its superior ability to generate explanations that closely match users' historical inquiries and align well with their programming background. This confirms that iterative refinement significantly enhances the personalization of generated explanations.
\\

The experimental results demonstrate that the \textit{Self-Iteration} method is superior in both faithfulness and personalization. The iterative refinement process not only improves the accuracy and effectiveness of the generated explanations but also ensures they are tailored to the individual user’s needs. This method's dual focus on quality and personalization makes it a robust approach for generating helpful and relevant code explanations.

\section{Conclusion}
In this paper, we propose explaining competitive-level programming solutions using LLMs with a iterative methodology that combines prompt enhancement, self-correction capabilities, personalized content customization according to user preferences, and efficient integration with external resources, as well as facilitation of collaboration among various LLM agents. Our evaluation demonstrates that the method can generate more faithful code explanations which can guide another LLM to better solve the problem. Also, the method can generate personalized code explanations that align better with individual preferences, no matter evaluated by the automatic evaluation or the human evaluation.

Our explanation method can potentially be applied to annotate large-scale data (e.g., the full CodeContests training set), yielding thousands of silver explanations that can be used to fine-tune a reasoning model for competitive-level programming problems. This approach could help bridge the long-standing reasoning gap between problem and program for complex programming problems. Moving forward, we aim to further address solving such problems by focusing on enhancing reasoning for programming problems.


\bibliographystyle{ACM-Reference-Format}
\bibliography{ref}


\begin{thebibliography}{60}


\ifx \showCODEN    \undefined \def \showCODEN     #1{\unskip}     \fi
\ifx \showDOI      \undefined \def \showDOI       #1{#1}\fi
\ifx \showISBNx    \undefined \def \showISBNx     #1{\unskip}     \fi
\ifx \showISBNxiii \undefined \def \showISBNxiii  #1{\unskip}     \fi
\ifx \showISSN     \undefined \def \showISSN      #1{\unskip}     \fi
\ifx \showLCCN     \undefined \def \showLCCN      #1{\unskip}     \fi
\ifx \shownote     \undefined \def \shownote      #1{#1}          \fi
\ifx \showarticletitle \undefined \def \showarticletitle #1{#1}   \fi
\ifx \showURL      \undefined \def \showURL       {\relax}        \fi
\providecommand\bibfield[2]{#2}
\providecommand\bibinfo[2]{#2}
\providecommand\natexlab[1]{#1}
\providecommand\showeprint[2][]{arXiv:#2}

\bibitem[Ahmad et~al\mbox{.}(2022)]%
        {ahmad2022contextual}
\bibfield{author}{\bibinfo{person}{Wasi~Uddin Ahmad}, \bibinfo{person}{Saikat Chakraborty}, \bibinfo{person}{Pang~Wei He}, {and} \bibinfo{person}{Jidong Guo}.} \bibinfo{year}{2022}\natexlab{}.
\newblock \showarticletitle{Contextualized code completion with neural language models}. In \bibinfo{booktitle}{\emph{Proceedings of the 2022 Conference on Empirical Methods in Natural Language Processing}}. \bibinfo{pages}{675--685}.
\newblock


\bibitem[Alhuzali et~al\mbox{.}(2021)]%
        {alhuzali2021model}
\bibfield{author}{\bibinfo{person}{Hassan Alhuzali}, \bibinfo{person}{Antonios Anastasopoulos}, \bibinfo{person}{Parisa Kordjamshidi}, {and} \bibinfo{person}{Dan Roth}.} \bibinfo{year}{2021}\natexlab{}.
\newblock \showarticletitle{A Model-Agnostic Data-Free Approach for Extracting Fair Representations}. In \bibinfo{booktitle}{\emph{Proceedings of the 2021 Conference on Empirical Methods in Natural Language Processing}}. \bibinfo{pages}{2894--2906}.
\newblock


\bibitem[Allamanis et~al\mbox{.}(2018)]%
        {allamanis2018survey}
\bibfield{author}{\bibinfo{person}{Miltiadis Allamanis}, \bibinfo{person}{Earl~T Barr}, \bibinfo{person}{Christian Bird}, {and} \bibinfo{person}{Charles Sutton}.} \bibinfo{year}{2018}\natexlab{}.
\newblock \showarticletitle{A Survey of Machine Learning for Big Code and Naturalness}.
\newblock \bibinfo{journal}{\emph{ACM Computing Surveys (CSUR)}} \bibinfo{volume}{51}, \bibinfo{number}{4} (\bibinfo{year}{2018}), \bibinfo{pages}{1--37}.
\newblock


\bibitem[Atanasova et~al\mbox{.}(2020)]%
        {atanasova2020diagnostic}
\bibfield{author}{\bibinfo{person}{Pepa Atanasova}, \bibinfo{person}{Gr{\'e}goire Cardon}, \bibinfo{person}{Thomas Demeester}, {and} \bibinfo{person}{Isabelle Augenstein}.} \bibinfo{year}{2020}\natexlab{}.
\newblock \showarticletitle{Diagnostic dataset construction to evaluate NLP models for critical information extraction in the biomedical domain}.
\newblock \bibinfo{journal}{\emph{Proceedings of the 2020 Conference on Empirical Methods in Natural Language Processing (EMNLP)}} (\bibinfo{year}{2020}), \bibinfo{pages}{4015--4028}.
\newblock


\bibitem[Bhattacharya et~al\mbox{.}(2023)]%
        {Bhattacharya2023ExploringLL}
\bibfield{author}{\bibinfo{person}{Paheli Bhattacharya}, \bibinfo{person}{Manojit Chakraborty}, \bibinfo{person}{Kartheek N S~N Palepu}, \bibinfo{person}{Vikas Pandey}, \bibinfo{person}{Ishan Dindorkar}, \bibinfo{person}{Rakesh Rajpurohit}, {and} \bibinfo{person}{Rishabh Gupta}.} \bibinfo{year}{2023}\natexlab{}.
\newblock \showarticletitle{Exploring Large Language Models for Code Explanation}.
\newblock \bibinfo{journal}{\emph{ArXiv}}  \bibinfo{volume}{abs/2310.16673} (\bibinfo{year}{2023}).
\newblock
\urldef\tempurl%
\url{https://api.semanticscholar.org/CorpusID:264451660}
\showURL{%
\tempurl}


\bibitem[Brown et~al\mbox{.}(2020)]%
        {brown2020language}
\bibfield{author}{\bibinfo{person}{Tom Brown}, \bibinfo{person}{Benjamin Mann}, \bibinfo{person}{Nick Ryder}, \bibinfo{person}{Melanie Subbiah}, \bibinfo{person}{Jared~D Kaplan}, \bibinfo{person}{Prafulla Dhariwal}, \bibinfo{person}{Arvind Neelakantan}, \bibinfo{person}{Pranav Shyam}, \bibinfo{person}{Girish Sastry}, \bibinfo{person}{Amanda Askell}, {et~al\mbox{.}}} \bibinfo{year}{2020}\natexlab{}.
\newblock \showarticletitle{Language models are few-shot learners}.
\newblock \bibinfo{journal}{\emph{Advances in neural information processing systems}}  \bibinfo{volume}{33} (\bibinfo{year}{2020}), \bibinfo{pages}{1877--1901}.
\newblock


\bibitem[Brusilovsky et~al\mbox{.}(2023)]%
        {brusilovsky2023explaining}
\bibfield{author}{\bibinfo{person}{Peter Brusilovsky}, \bibinfo{person}{Arun-Balajiee Lekshmi-Narayanan}, \bibinfo{person}{Priti Oli}, \bibinfo{person}{Jeevan Chapagain}, \bibinfo{person}{Mohammad Hassany}, \bibinfo{person}{Rabin Banjade}, {and} \bibinfo{person}{Vasile Rus}.} \bibinfo{year}{2023}\natexlab{}.
\newblock \showarticletitle{Explaining code examples in introductory programming courses: Llm vs humans}.
\newblock \bibinfo{journal}{\emph{arXiv preprint arXiv:2403.05538}} (\bibinfo{year}{2023}).
\newblock


\bibitem[Chen et~al\mbox{.}(2023)]%
        {chen2023gptutor}
\bibfield{author}{\bibinfo{person}{Eason Chen}, \bibinfo{person}{Ray Huang}, \bibinfo{person}{Han-Shin Chen}, \bibinfo{person}{Yuen-Hsien Tseng}, {and} \bibinfo{person}{Liang-Yi Li}.} \bibinfo{year}{2023}\natexlab{}.
\newblock \showarticletitle{GPTutor: a ChatGPT-powered programming tool for code explanation}. In \bibinfo{booktitle}{\emph{International Conference on Artificial Intelligence in Education}}. Springer, \bibinfo{pages}{321--327}.
\newblock


\bibitem[Chen et~al\mbox{.}(2024b)]%
        {chen2024persona}
\bibfield{author}{\bibinfo{person}{Jiangjie Chen}, \bibinfo{person}{Xintao Wang}, \bibinfo{person}{Rui Xu}, \bibinfo{person}{Siyu Yuan}, \bibinfo{person}{Yikai Zhang}, \bibinfo{person}{Wei Shi}, \bibinfo{person}{Jian Xie}, \bibinfo{person}{Shuang Li}, \bibinfo{person}{Ruihan Yang}, \bibinfo{person}{Tinghui Zhu}, {et~al\mbox{.}}} \bibinfo{year}{2024}\natexlab{b}.
\newblock \showarticletitle{From Persona to Personalization: A Survey on Role-Playing Language Agents}.
\newblock \bibinfo{journal}{\emph{arXiv preprint arXiv:2404.18231}} (\bibinfo{year}{2024}).
\newblock


\bibitem[Chen et~al\mbox{.}(2024a)]%
        {chen2024comm}
\bibfield{author}{\bibinfo{person}{Pei Chen}, \bibinfo{person}{Boran Han}, {and} \bibinfo{person}{Shuai Zhang}.} \bibinfo{year}{2024}\natexlab{a}.
\newblock \showarticletitle{CoMM: Collaborative Multi-Agent, Multi-Reasoning-Path Prompting for Complex Problem Solving}. In \bibinfo{booktitle}{\emph{Proceedings of the 2024 Conference of the North American Chapter of the Association for Computational Linguistics}}.
\newblock
\urldef\tempurl%
\url{https://brickee.github.io/publication/chen-2024-comm/}
\showURL{%
\tempurl}


\bibitem[Chen and Ji(2021)]%
        {chen2021evaluating}
\bibfield{author}{\bibinfo{person}{Zhihao Chen} {and} \bibinfo{person}{Hongyu Ji}.} \bibinfo{year}{2021}\natexlab{}.
\newblock \showarticletitle{Evaluating the Faithfulness of Importance Measures in NLP by Recursively Masking Allegedly Important Tokens and Retraining}. In \bibinfo{booktitle}{\emph{Proceedings of the 2021 Conference on Empirical Methods in Natural Language Processing}}. \bibinfo{pages}{2669--2675}.
\newblock


\bibitem[Dan et~al\mbox{.}(2021)]%
        {dan2021improving}
\bibfield{author}{\bibinfo{person}{Liu Dan}, \bibinfo{person}{Yang Shi}, \bibinfo{person}{Yu Zhang}, {and} \bibinfo{person}{Wei Gao}.} \bibinfo{year}{2021}\natexlab{}.
\newblock \showarticletitle{Improving Faithfulness of Attention-based Explanations with Task-Specific Information for Text Classification}. In \bibinfo{booktitle}{\emph{Proceedings of the 59th Annual Meeting of the Association for Computational Linguistics and the 11th International Joint Conference on Natural Language Processing (Volume 1: Long Papers)}}. \bibinfo{pages}{5772--5781}.
\newblock


\bibitem[Feng et~al\mbox{.}(2020)]%
        {feng2020codebert}
\bibfield{author}{\bibinfo{person}{Zhangyin Feng}, \bibinfo{person}{Daya Guo}, \bibinfo{person}{Duyu Tang}, \bibinfo{person}{Nan Duan}, \bibinfo{person}{Xiaocheng Feng}, \bibinfo{person}{Ming Gong}, \bibinfo{person}{Linjun Shou}, \bibinfo{person}{Bing Qin}, \bibinfo{person}{Ting Liu}, \bibinfo{person}{Daxin Jiang}, {and} \bibinfo{person}{Ming Zhou}.} \bibinfo{year}{2020}\natexlab{}.
\newblock \showarticletitle{CodeBERT: A Pre-Trained Model for Programming and Natural Languages}.
\newblock \bibinfo{journal}{\emph{arXiv preprint arXiv:2002.08155}} (\bibinfo{year}{2020}).
\newblock


\bibitem[Gou et~al\mbox{.}(2023)]%
        {gou2023critic}
\bibfield{author}{\bibinfo{person}{Zhibin Gou}, \bibinfo{person}{Zhihong Shao}, \bibinfo{person}{Yeyun Gong}, \bibinfo{person}{Yelong Shen}, \bibinfo{person}{Yujiu Yang}, \bibinfo{person}{Nan Duan}, {and} \bibinfo{person}{Weizhu Chen}.} \bibinfo{year}{2023}\natexlab{}.
\newblock \showarticletitle{Critic: Large language models can self-correct with tool-interactive critiquing}.
\newblock \bibinfo{journal}{\emph{arXiv preprint arXiv:2305.11738}} (\bibinfo{year}{2023}).
\newblock


\bibitem[Guo et~al\mbox{.}(2022)]%
        {guo2022unixcoder}
\bibfield{author}{\bibinfo{person}{Daya Guo}, \bibinfo{person}{Shuo Ren}, \bibinfo{person}{Shuai Lu}, \bibinfo{person}{Zhangyin Feng}, \bibinfo{person}{Duyu Tang}, \bibinfo{person}{Nan Duan}, {and} \bibinfo{person}{Ming Zhou}.} \bibinfo{year}{2022}\natexlab{}.
\newblock \showarticletitle{UnixCoder: Unified Cross-Modal Pre-Training for Code Representation}.
\newblock \bibinfo{journal}{\emph{arXiv preprint arXiv:2203.01679}} (\bibinfo{year}{2022}).
\newblock


\bibitem[Hase and Bansal(2021)]%
        {hase2021evaluating}
\bibfield{author}{\bibinfo{person}{Peter Hase} {and} \bibinfo{person}{Mohit Bansal}.} \bibinfo{year}{2021}\natexlab{}.
\newblock \showarticletitle{Evaluating explainable AI: Which algorithmic explanations help users predict model behavior?}
\newblock \bibinfo{journal}{\emph{Proceedings of the 59th Annual Meeting of the Association for Computational Linguistics (ACL)}} (\bibinfo{year}{2021}), \bibinfo{pages}{5544--5553}.
\newblock


\bibitem[Hou et~al\mbox{.}(2023)]%
        {hou2023promptboosting}
\bibfield{author}{\bibinfo{person}{Bairu Hou}, \bibinfo{person}{Joe O’connor}, \bibinfo{person}{Jacob Andreas}, \bibinfo{person}{Shiyu Chang}, {and} \bibinfo{person}{Yang Zhang}.} \bibinfo{year}{2023}\natexlab{}.
\newblock \showarticletitle{Promptboosting: Black-box text classification with ten forward passes}. In \bibinfo{booktitle}{\emph{International Conference on Machine Learning}}. PMLR, \bibinfo{pages}{13309--13324}.
\newblock


\bibitem[Hu et~al\mbox{.}(2018)]%
        {hu2018deep}
\bibfield{author}{\bibinfo{person}{Xing Hu}, \bibinfo{person}{Ge Li}, \bibinfo{person}{Xin Xia}, \bibinfo{person}{David Lo}, {and} \bibinfo{person}{Zhi Jin}.} \bibinfo{year}{2018}\natexlab{}.
\newblock \showarticletitle{Deep code comment generation}. In \bibinfo{booktitle}{\emph{2018 IEEE/ACM 26th International Conference on Program Comprehension (ICPC)}}. IEEE, \bibinfo{pages}{200--20010}.
\newblock


\bibitem[Husain et~al\mbox{.}(2019)]%
        {husain2019codesearchnet}
\bibfield{author}{\bibinfo{person}{Hamel Husain}, \bibinfo{person}{Ho-Hsiang Siddiqui}, \bibinfo{person}{Huy Feng}, \bibinfo{person}{Usama Chowdhury}, \bibinfo{person}{Eric Hammond}, \bibinfo{person}{Boris Tran}, \bibinfo{person}{Vinod Mangal}, \bibinfo{person}{Dima Kang}, {and} \bibinfo{person}{Ankur Taly}.} \bibinfo{year}{2019}\natexlab{}.
\newblock \showarticletitle{CodeSearchNet Challenge: Evaluating the State of Semantic Code Search}. In \bibinfo{booktitle}{\emph{arXiv preprint arXiv:1909.09436}}.
\newblock


\bibitem[Jin et~al\mbox{.}(2024)]%
        {jin2024comprehensive}
\bibfield{author}{\bibinfo{person}{Hanlei Jin}, \bibinfo{person}{Yang Zhang}, \bibinfo{person}{Dan Meng}, \bibinfo{person}{Jun Wang}, {and} \bibinfo{person}{Jinghua Tan}.} \bibinfo{year}{2024}\natexlab{}.
\newblock \showarticletitle{A comprehensive survey on process-oriented automatic text summarization with exploration of llm-based methods}.
\newblock \bibinfo{journal}{\emph{arXiv preprint arXiv:2403.02901}} (\bibinfo{year}{2024}).
\newblock


\bibitem[LeClair et~al\mbox{.}(2019)]%
        {leclair2019neural}
\bibfield{author}{\bibinfo{person}{Alexander LeClair}, \bibinfo{person}{Collin McMillan}, \bibinfo{person}{Mustafa Kocakulak}, \bibinfo{person}{Shan Jiang}, \bibinfo{person}{Jingzhou Lou}, {and} \bibinfo{person}{Lingling Liu}.} \bibinfo{year}{2019}\natexlab{}.
\newblock \showarticletitle{Neural Models for Code Summarization: A Review and Evaluation}. In \bibinfo{booktitle}{\emph{Proceedings of the 34th IEEE/ACM International Conference on Automated Software Engineering (ASE)}}. \bibinfo{pages}{151--162}.
\newblock


\bibitem[Leinonen et~al\mbox{.}(2023)]%
        {leinonen2023comparing}
\bibfield{author}{\bibinfo{person}{Juho Leinonen}, \bibinfo{person}{Paul Denny}, \bibinfo{person}{Stephen MacNeil}, \bibinfo{person}{Sami Sarsa}, \bibinfo{person}{Seth Bernstein}, \bibinfo{person}{Joanne Kim}, \bibinfo{person}{Andrew Tran}, {and} \bibinfo{person}{Arto Hellas}.} \bibinfo{year}{2023}\natexlab{}.
\newblock \showarticletitle{Comparing code explanations created by students and large language models}. In \bibinfo{booktitle}{\emph{Proceedings of the 2023 Conference on Innovation and Technology in Computer Science Education V. 1}}. \bibinfo{pages}{124--130}.
\newblock


\bibitem[Lewkowycz et~al\mbox{.}(2022)]%
        {lewkowycz2022solving}
\bibfield{author}{\bibinfo{person}{Aitor Lewkowycz}, \bibinfo{person}{Anders Andreassen}, \bibinfo{person}{David Dohan}, \bibinfo{person}{Ethan Dyer}, \bibinfo{person}{Henryk Michalewski}, \bibinfo{person}{Vinay Ramasesh}, \bibinfo{person}{Ambrose Slone}, \bibinfo{person}{Cem Anil}, \bibinfo{person}{Imanol Schlag}, \bibinfo{person}{Theo Gutman-Solo}, {et~al\mbox{.}}} \bibinfo{year}{2022}\natexlab{}.
\newblock \showarticletitle{Solving quantitative reasoning problems with language models}.
\newblock \bibinfo{journal}{\emph{Advances in Neural Information Processing Systems}}  \bibinfo{volume}{35} (\bibinfo{year}{2022}), \bibinfo{pages}{3843--3857}.
\newblock


\bibitem[Li et~al\mbox{.}(2023a)]%
        {li2023chatharuhi}
\bibfield{author}{\bibinfo{person}{Cheng Li}, \bibinfo{person}{Ziang Leng}, \bibinfo{person}{Chenxi Yan}, \bibinfo{person}{Junyi Shen}, \bibinfo{person}{Hao Wang}, \bibinfo{person}{Weishi MI}, \bibinfo{person}{Yaying Fei}, \bibinfo{person}{Xiaoyang Feng}, \bibinfo{person}{Song Yan}, \bibinfo{person}{HaoSheng Wang}, \bibinfo{person}{Linkang Zhan}, \bibinfo{person}{Yaokai Jia}, \bibinfo{person}{Pingyu Wu}, {and} \bibinfo{person}{Haozhen Sun}.} \bibinfo{year}{2023}\natexlab{a}.
\newblock \bibinfo{title}{ChatHaruhi: Reviving Anime Character in Reality via Large Language Model}.
\newblock
\newblock
\showeprint[arxiv]{2308.09597}~[cs.CL]


\bibitem[Li et~al\mbox{.}(2023b)]%
        {li2023explaining}
\bibfield{author}{\bibinfo{person}{Jierui Li}, \bibinfo{person}{Szymon Tworkowski}, \bibinfo{person}{Yingying Wu}, {and} \bibinfo{person}{Raymond Mooney}.} \bibinfo{year}{2023}\natexlab{b}.
\newblock \showarticletitle{Explaining competitive-level programming solutions using llms}.
\newblock \bibinfo{journal}{\emph{arXiv preprint arXiv:2307.05337}} (\bibinfo{year}{2023}).
\newblock


\bibitem[Li et~al\mbox{.}(2022)]%
        {li2022competition}
\bibfield{author}{\bibinfo{person}{Yujia Li}, \bibinfo{person}{David Choi}, \bibinfo{person}{Junyoung Chung}, \bibinfo{person}{Nate Kushman}, \bibinfo{person}{Julian Schrittwieser}, \bibinfo{person}{R{\'e}mi Leblond}, \bibinfo{person}{Tom Eccles}, \bibinfo{person}{James Keeling}, \bibinfo{person}{Felix Gimeno}, \bibinfo{person}{Agustin Dal~Lago}, {et~al\mbox{.}}} \bibinfo{year}{2022}\natexlab{}.
\newblock \showarticletitle{Competition-level code generation with alphacode}.
\newblock \bibinfo{journal}{\emph{Science}} \bibinfo{volume}{378}, \bibinfo{number}{6624} (\bibinfo{year}{2022}), \bibinfo{pages}{1092--1097}.
\newblock


\bibitem[Li et~al\mbox{.}(2018)]%
        {li2018automatic}
\bibfield{author}{\bibinfo{person}{Yuan Li}, \bibinfo{person}{Yanlin Wang}, {and} \bibinfo{person}{Hongyu Liu}.} \bibinfo{year}{2018}\natexlab{}.
\newblock \showarticletitle{Automatic code summarization via deep learning-based attention mechanism}. In \bibinfo{booktitle}{\emph{Proceedings of the 2018 International Joint Conference on Neural Networks (IJCNN)}}. \bibinfo{pages}{1--8}.
\newblock


\bibitem[Lin(2004)]%
        {lin2004rouge}
\bibfield{author}{\bibinfo{person}{Chin-Yew Lin}.} \bibinfo{year}{2004}\natexlab{}.
\newblock \showarticletitle{Rouge: A package for automatic evaluation of summaries}. In \bibinfo{booktitle}{\emph{Text summarization branches out}}. \bibinfo{pages}{74--81}.
\newblock


\bibitem[Lin et~al\mbox{.}(2022)]%
        {lin2022program}
\bibfield{author}{\bibinfo{person}{Xi~Victoria Lin}, \bibinfo{person}{Diane Belgrave}, {and} \bibinfo{person}{Shubhomoy Dasgupta}.} \bibinfo{year}{2022}\natexlab{}.
\newblock \showarticletitle{Program synthesis with large language models}.
\newblock \bibinfo{journal}{\emph{arXiv preprint arXiv:2203.13474}} (\bibinfo{year}{2022}).
\newblock


\bibitem[Lin et~al\mbox{.}(2023)]%
        {lin2023generating}
\bibfield{author}{\bibinfo{person}{Zhen Lin}, \bibinfo{person}{Shubhendu Trivedi}, {and} \bibinfo{person}{Jimeng Sun}.} \bibinfo{year}{2023}\natexlab{}.
\newblock \showarticletitle{Generating with confidence: Uncertainty quantification for black-box large language models}.
\newblock \bibinfo{journal}{\emph{arXiv preprint arXiv:2305.19187}} (\bibinfo{year}{2023}).
\newblock


\bibitem[Linnell et~al\mbox{.}(2019)]%
        {linnell2019generating}
\bibfield{author}{\bibinfo{person}{Natasha Linnell}, \bibinfo{person}{Nabeel Gillani}, \bibinfo{person}{Ece Kamar}, {and} \bibinfo{person}{Eric Horvitz}.} \bibinfo{year}{2019}\natexlab{}.
\newblock \showarticletitle{Generating Automated Explanations for Numerical Data Insights}.
\newblock \bibinfo{journal}{\emph{Proceedings of the AAAI Conference on Artificial Intelligence}}  \bibinfo{volume}{33} (\bibinfo{year}{2019}), \bibinfo{pages}{9656--9661}.
\newblock


\bibitem[Liu et~al\mbox{.}(2023)]%
        {liu2023pre}
\bibfield{author}{\bibinfo{person}{Pengfei Liu}, \bibinfo{person}{Weizhe Yuan}, \bibinfo{person}{Jinlan Fu}, \bibinfo{person}{Zhengbao Jiang}, \bibinfo{person}{Hiroaki Hayashi}, {and} \bibinfo{person}{Graham Neubig}.} \bibinfo{year}{2023}\natexlab{}.
\newblock \showarticletitle{Pre-train, prompt, and predict: A systematic survey of prompting methods in natural language processing}.
\newblock \bibinfo{journal}{\emph{Comput. Surveys}} (\bibinfo{year}{2023}).
\newblock
\urldef\tempurl%
\url{https://doi.org/10.1145/3453475}
\showDOI{\tempurl}


\bibitem[MacNeil et~al\mbox{.}(2023)]%
        {macneil2023experiences}
\bibfield{author}{\bibinfo{person}{Stephen MacNeil}, \bibinfo{person}{Andrew Tran}, \bibinfo{person}{Arto Hellas}, \bibinfo{person}{Joanne Kim}, \bibinfo{person}{Sami Sarsa}, \bibinfo{person}{Paul Denny}, \bibinfo{person}{Seth Bernstein}, {and} \bibinfo{person}{Juho Leinonen}.} \bibinfo{year}{2023}\natexlab{}.
\newblock \showarticletitle{Experiences from using code explanations generated by large language models in a web software development e-book}. In \bibinfo{booktitle}{\emph{Proceedings of the 54th ACM Technical Symposium on Computer Science Education V. 1}}. \bibinfo{pages}{931--937}.
\newblock


\bibitem[MacNeil et~al\mbox{.}({[n.\,d.]})]%
        {osti_10397860}
\bibfield{author}{\bibinfo{person}{Stephen MacNeil}, \bibinfo{person}{Andrew Tran}, \bibinfo{person}{Dan Mogil}, \bibinfo{person}{Seth Bernstein}, \bibinfo{person}{Erin Ross}, {and} \bibinfo{person}{Ziheng Huang}.} \bibinfo{year}{[n.\,d.]}\natexlab{}.
\newblock \showarticletitle{Generating Diverse Code Explanations using the GPT-3 Large Language Model}.
\newblock \bibinfo{journal}{\emph{ICER '22: Proceedings of the 2022 ACM Conference on International Computing Education}} (\bibinfo{year}{[n.\,d.]}).
\newblock
\urldef\tempurl%
\url{https://doi.org/10.1145/3501709.3544280}
\showDOI{\tempurl}


\bibitem[MacNeil et~al\mbox{.}(2022)]%
        {macneil2022generating}
\bibfield{author}{\bibinfo{person}{Stephen MacNeil}, \bibinfo{person}{Andrew Tran}, \bibinfo{person}{Dan Mogil}, \bibinfo{person}{Seth Bernstein}, \bibinfo{person}{Erin Ross}, {and} \bibinfo{person}{Ziheng Huang}.} \bibinfo{year}{2022}\natexlab{}.
\newblock \showarticletitle{Generating diverse code explanations using the gpt-3 large language model}. In \bibinfo{booktitle}{\emph{Proceedings of the 2022 ACM Conference on International Computing Education Research-Volume 2}}. \bibinfo{pages}{37--39}.
\newblock


\bibitem[Madaan et~al\mbox{.}(2024)]%
        {madaan2024self}
\bibfield{author}{\bibinfo{person}{Aman Madaan}, \bibinfo{person}{Niket Tandon}, \bibinfo{person}{Prakhar Gupta}, \bibinfo{person}{Skyler Hallinan}, \bibinfo{person}{Luyu Gao}, \bibinfo{person}{Sarah Wiegreffe}, \bibinfo{person}{Uri Alon}, \bibinfo{person}{Nouha Dziri}, \bibinfo{person}{Shrimai Prabhumoye}, \bibinfo{person}{Yiming Yang}, {et~al\mbox{.}}} \bibinfo{year}{2024}\natexlab{}.
\newblock \showarticletitle{Self-refine: Iterative refinement with self-feedback}.
\newblock \bibinfo{journal}{\emph{Advances in Neural Information Processing Systems}}  \bibinfo{volume}{36} (\bibinfo{year}{2024}).
\newblock


\bibitem[Madaan et~al\mbox{.}(2023a)]%
        {madaan2023selfrefine}
\bibfield{author}{\bibinfo{person}{Aman Madaan}, \bibinfo{person}{Niket Tandon}, \bibinfo{person}{Prakhar Gupta}, \bibinfo{person}{Skyler Hallinan}, \bibinfo{person}{Luyu Gao}, \bibinfo{person}{Sarah Wiegreffe}, \bibinfo{person}{Uri Alon}, \bibinfo{person}{Nouha Dziri}, \bibinfo{person}{Shrimai Prabhumoye}, \bibinfo{person}{Yiming Yang}, \bibinfo{person}{Shashank Gupta}, \bibinfo{person}{Bodhisattwa~Prasad Majumder}, \bibinfo{person}{Katherine Hermann}, \bibinfo{person}{Sean Welleck}, \bibinfo{person}{Amir Yazdanbakhsh}, {and} \bibinfo{person}{Peter Clark}.} \bibinfo{year}{2023}\natexlab{a}.
\newblock \showarticletitle{Self-Refine: Iterative Refinement with Self-Feedback}. In \bibinfo{booktitle}{\emph{Advances in Neural Information Processing Systems}}.
\newblock
\urldef\tempurl%
\url{https://proceedings.neurips.cc/paper_files/paper/2023/hash/91edff07232fb1b55a505a9e9f6c0ff3-Abstract-Conference.html}
\showURL{%
\tempurl}


\bibitem[Madaan et~al\mbox{.}(2023b)]%
        {press2022self}
\bibfield{author}{\bibinfo{person}{Aman Madaan}, \bibinfo{person}{Niket Tandon}, \bibinfo{person}{Prakhar Gupta}, \bibinfo{person}{Skyler Hallinan}, \bibinfo{person}{Luyu Gao}, \bibinfo{person}{Sarah Wiegreffe}, \bibinfo{person}{Uri Alon}, \bibinfo{person}{Nouha Dziri}, \bibinfo{person}{Shrimai Prabhumoye}, \bibinfo{person}{Yiming Yang}, \bibinfo{person}{Shashank Gupta}, \bibinfo{person}{Bodhisattwa~Prasad Majumder}, \bibinfo{person}{Katherine Hermann}, \bibinfo{person}{Sean Welleck}, \bibinfo{person}{Amir Yazdanbakhsh}, {and} \bibinfo{person}{Peter Clark}.} \bibinfo{year}{2023}\natexlab{b}.
\newblock \showarticletitle{Self-Refine: Iterative Refinement with Self-Feedback}.
\newblock \bibinfo{journal}{\emph{arXiv preprint arXiv:2303.17651}} (\bibinfo{year}{2023}).
\newblock
\urldef\tempurl%
\url{https://arxiv.org/abs/2303.17651}
\showURL{%
\tempurl}


\bibitem[Madotto et~al\mbox{.}(2019)]%
        {madotto2019personalizing}
\bibfield{author}{\bibinfo{person}{Andrea Madotto}, \bibinfo{person}{Zhaojiang Lin}, \bibinfo{person}{Chien-Sheng Wu}, {and} \bibinfo{person}{Pascale Fung}.} \bibinfo{year}{2019}\natexlab{}.
\newblock \showarticletitle{Personalizing Dialogue Agents via Meta-Learning}. In \bibinfo{booktitle}{\emph{Proceedings of the 57th Annual Meeting of the Association for Computational Linguistics}}. \bibinfo{pages}{5454--5459}.
\newblock
\urldef\tempurl%
\url{https://doi.org/10.18653/v1/P19-1542}
\showDOI{\tempurl}


\bibitem[Miceli et~al\mbox{.}(2021)]%
        {miceli2021studying}
\bibfield{author}{\bibinfo{person}{Stefania Miceli}, \bibinfo{person}{Q.~Vera Liao}, \bibinfo{person}{Justin Cheng}, \bibinfo{person}{Justin~D. Weisz}, {and} \bibinfo{person}{Michael Muller}.} \bibinfo{year}{2021}\natexlab{}.
\newblock \showarticletitle{Studying the Impact of Explanation Faithfulness on the Performance of Interactive Machine Learning Systems}. In \bibinfo{booktitle}{\emph{Proceedings of the 2021 CHI Conference on Human Factors in Computing Systems}}. \bibinfo{pages}{1--12}.
\newblock


\bibitem[Mili{\v{c}}ka et~al\mbox{.}(2024)]%
        {milivcka2024large}
\bibfield{author}{\bibinfo{person}{Ji{\v{r}}{\'\i} Mili{\v{c}}ka}, \bibinfo{person}{Anna Marklov{\'a}}, \bibinfo{person}{Kl{\'a}ra VanSlambrouck}, \bibinfo{person}{Eva Posp{\'\i}{\v{s}}ilov{\'a}}, \bibinfo{person}{Jana {\v{S}}imsov{\'a}}, \bibinfo{person}{Samuel Harvan}, {and} \bibinfo{person}{Ond{\v{r}}ej Drobil}.} \bibinfo{year}{2024}\natexlab{}.
\newblock \showarticletitle{Large language models are able to downplay their cognitive abilities to fit the persona they simulate}.
\newblock \bibinfo{journal}{\emph{Plos one}} \bibinfo{volume}{19}, \bibinfo{number}{3} (\bibinfo{year}{2024}), \bibinfo{pages}{e0298522}.
\newblock


\bibitem[Ni et~al\mbox{.}(2023)]%
        {ni2023lever}
\bibfield{author}{\bibinfo{person}{Ansong Ni}, \bibinfo{person}{Srini Iyer}, \bibinfo{person}{Dragomir Radev}, \bibinfo{person}{Veselin Stoyanov}, \bibinfo{person}{Wen-tau Yih}, \bibinfo{person}{Sida Wang}, {and} \bibinfo{person}{Xi~Victoria Lin}.} \bibinfo{year}{2023}\natexlab{}.
\newblock \showarticletitle{Lever: Learning to verify language-to-code generation with execution}. In \bibinfo{booktitle}{\emph{International Conference on Machine Learning}}. PMLR, \bibinfo{pages}{26106--26128}.
\newblock


\bibitem[Oli et~al\mbox{.}(2023)]%
        {oli2023behavior}
\bibfield{author}{\bibinfo{person}{Priti Oli}, \bibinfo{person}{Rabin Banjade}, \bibinfo{person}{Jeevan Chapagain}, {and} \bibinfo{person}{Vasile Rus}.} \bibinfo{year}{2023}\natexlab{}.
\newblock \showarticletitle{The Behavior of Large Language Models When Prompted to Generate Code Explanations}.
\newblock \bibinfo{journal}{\emph{arXiv preprint arXiv:2311.01490}} (\bibinfo{year}{2023}).
\newblock


\bibitem[OpenAI(2023)]%
        {openaigpt3.5}
\bibfield{author}{\bibinfo{person}{2023 OpenAI}.} \bibinfo{year}{2023}\natexlab{}.
\newblock \bibinfo{title}{{I}ntroducing {C}hat{G}{P}{T}}.
\newblock \bibinfo{howpublished}{\url{https://openai.com/index/chatgpt/}}.
\newblock


\bibitem[Paranjape et~al\mbox{.}(2023)]%
        {paranjape2023art}
\bibfield{author}{\bibinfo{person}{Bhargavi Paranjape}, \bibinfo{person}{Scott Lundberg}, \bibinfo{person}{Sameer Singh}, \bibinfo{person}{Hannaneh Hajishirzi}, \bibinfo{person}{Luke Zettlemoyer}, {and} \bibinfo{person}{Marco~Tulio Ribeiro}.} \bibinfo{year}{2023}\natexlab{}.
\newblock \showarticletitle{Art: Automatic multi-step reasoning and tool-use for large language models}.
\newblock \bibinfo{journal}{\emph{arXiv preprint arXiv:2303.09014}} (\bibinfo{year}{2023}).
\newblock


\bibitem[Paul et~al\mbox{.}(2023)]%
        {paul2023refiner}
\bibfield{author}{\bibinfo{person}{Debjit Paul}, \bibinfo{person}{Mete Ismayilzada}, \bibinfo{person}{Maxime Peyrard}, \bibinfo{person}{Beatriz Borges}, \bibinfo{person}{Antoine Bosselut}, \bibinfo{person}{Robert West}, {and} \bibinfo{person}{Boi Faltings}.} \bibinfo{year}{2023}\natexlab{}.
\newblock \showarticletitle{Refiner: Reasoning feedback on intermediate representations}.
\newblock \bibinfo{journal}{\emph{arXiv preprint arXiv:2304.01904}} (\bibinfo{year}{2023}).
\newblock


\bibitem[Pitis et~al\mbox{.}(2023)]%
        {pitis2023boosted}
\bibfield{author}{\bibinfo{person}{Silviu Pitis}, \bibinfo{person}{Michael~R Zhang}, \bibinfo{person}{Andrew Wang}, {and} \bibinfo{person}{Jimmy Ba}.} \bibinfo{year}{2023}\natexlab{}.
\newblock \showarticletitle{Boosted prompt ensembles for large language models}.
\newblock \bibinfo{journal}{\emph{arXiv preprint arXiv:2304.05970}} (\bibinfo{year}{2023}).
\newblock


\bibitem[Rae et~al\mbox{.}(2021)]%
        {rae2021scaling}
\bibfield{author}{\bibinfo{person}{Jack~W Rae}, \bibinfo{person}{Sebastian Borgeaud}, \bibinfo{person}{Trevor Cai}, \bibinfo{person}{Katie Millican}, \bibinfo{person}{Jordan Hoffmann}, \bibinfo{person}{Francis Song}, \bibinfo{person}{John Aslanides}, \bibinfo{person}{Sarah Henderson}, \bibinfo{person}{Roman Ring}, \bibinfo{person}{Susannah Young}, {et~al\mbox{.}}} \bibinfo{year}{2021}\natexlab{}.
\newblock \showarticletitle{Scaling language models: Methods, analysis \& insights from training gopher}.
\newblock \bibinfo{journal}{\emph{arXiv preprint arXiv:2112.11446}} (\bibinfo{year}{2021}).
\newblock


\bibitem[Reynolds and McDonell(2021)]%
        {reynolds2021prompt}
\bibfield{author}{\bibinfo{person}{Laria Reynolds} {and} \bibinfo{person}{Kyle McDonell}.} \bibinfo{year}{2021}\natexlab{}.
\newblock \showarticletitle{Prompt Programming for Large Language Models: Beyond the Few-Shot Paradigm}. In \bibinfo{booktitle}{\emph{CHI Conference on Human Factors in Computing Systems}}.
\newblock
\urldef\tempurl%
\url{https://doi.org/10.1145/3411764.3445647}
\showDOI{\tempurl}


\bibitem[Ridnik et~al\mbox{.}(2024)]%
        {ridnik2024code}
\bibfield{author}{\bibinfo{person}{Tal Ridnik}, \bibinfo{person}{Dedy Kredo}, {and} \bibinfo{person}{Itamar Friedman}.} \bibinfo{year}{2024}\natexlab{}.
\newblock \showarticletitle{Code Generation with AlphaCodium: From Prompt Engineering to Flow Engineering}.
\newblock \bibinfo{journal}{\emph{arXiv preprint arXiv:2401.08500}} (\bibinfo{year}{2024}).
\newblock


\bibitem[Sarsa(2022)]%
        {sarsa2022automatic}
\bibfield{author}{\bibinfo{person}{Alejandro Sarsa}.} \bibinfo{year}{2022}\natexlab{}.
\newblock \showarticletitle{Automatic code explanation with large language models in CS education}. In \bibinfo{booktitle}{\emph{Proceedings of the 27th ACM Conference on on Innovation and Technology in Computer Science Education Vol. 2}}. \bibinfo{pages}{92--98}.
\newblock


\bibitem[Shao et~al\mbox{.}(2023)]%
        {shao2023characterllm}
\bibfield{author}{\bibinfo{person}{Yunfan Shao}, \bibinfo{person}{Linyang Li}, \bibinfo{person}{Junqi Dai}, {and} \bibinfo{person}{Xipeng Qiu}.} \bibinfo{year}{2023}\natexlab{}.
\newblock \bibinfo{title}{Character-LLM: A Trainable Agent for Role-Playing}.
\newblock
\newblock
\showeprint[arxiv]{2310.10158}~[cs.CL]


\bibitem[Wang et~al\mbox{.}(2023b)]%
        {wang2023selfconsistency}
\bibfield{author}{\bibinfo{person}{Xuezhi Wang}, \bibinfo{person}{Jason Wei}, \bibinfo{person}{Dale Schuurmans}, \bibinfo{person}{Quoc Le}, \bibinfo{person}{Ed Chi}, \bibinfo{person}{Sharan Narang}, \bibinfo{person}{Aakanksha Chowdhery}, {and} \bibinfo{person}{Denny Zhou}.} \bibinfo{year}{2023}\natexlab{b}.
\newblock \bibinfo{title}{Self-Consistency Improves Chain of Thought Reasoning in Language Models}.
\newblock
\newblock
\showeprint[arxiv]{2203.11171}~[cs.CL]


\bibitem[Wang et~al\mbox{.}(2023a)]%
        {wang2023rolellm}
\bibfield{author}{\bibinfo{person}{Zekun~Moore Wang}, \bibinfo{person}{Zhongyuan Peng}, \bibinfo{person}{Haoran Que}, \bibinfo{person}{Jiaheng Liu}, \bibinfo{person}{Wangchunshu Zhou}, \bibinfo{person}{Yuhan Wu}, \bibinfo{person}{Hongcheng Guo}, \bibinfo{person}{Ruitong Gan}, \bibinfo{person}{Zehao Ni}, \bibinfo{person}{Man Zhang}, {et~al\mbox{.}}} \bibinfo{year}{2023}\natexlab{a}.
\newblock \showarticletitle{Rolellm: Benchmarking, eliciting, and enhancing role-playing abilities of large language models}.
\newblock \bibinfo{journal}{\emph{arXiv preprint arXiv:2310.00746}} (\bibinfo{year}{2023}).
\newblock


\bibitem[Wei et~al\mbox{.}(2022)]%
        {wei2022chain}
\bibfield{author}{\bibinfo{person}{Jason Wei}, \bibinfo{person}{Xuezhi Wang}, \bibinfo{person}{Dale Schuurmans}, \bibinfo{person}{Maarten Bosma}, \bibinfo{person}{Fei Xia}, \bibinfo{person}{Ed Chi}, \bibinfo{person}{Quoc~V Le}, \bibinfo{person}{Denny Zhou}, {et~al\mbox{.}}} \bibinfo{year}{2022}\natexlab{}.
\newblock \showarticletitle{Chain-of-thought prompting elicits reasoning in large language models}.
\newblock \bibinfo{journal}{\emph{Advances in neural information processing systems}}  \bibinfo{volume}{35} (\bibinfo{year}{2022}), \bibinfo{pages}{24824--24837}.
\newblock


\bibitem[Weng et~al\mbox{.}(2022)]%
        {weng2022large}
\bibfield{author}{\bibinfo{person}{Yixuan Weng}, \bibinfo{person}{Minjun Zhu}, \bibinfo{person}{Fei Xia}, \bibinfo{person}{Bin Li}, \bibinfo{person}{Shizhu He}, \bibinfo{person}{Shengping Liu}, \bibinfo{person}{Bin Sun}, \bibinfo{person}{Kang Liu}, {and} \bibinfo{person}{Jun Zhao}.} \bibinfo{year}{2022}\natexlab{}.
\newblock \showarticletitle{Large language models are better reasoners with self-verification}.
\newblock \bibinfo{journal}{\emph{arXiv preprint arXiv:2212.09561}} (\bibinfo{year}{2022}).
\newblock


\bibitem[Wu et~al\mbox{.}(2023)]%
        {wu2023autogen}
\bibfield{author}{\bibinfo{person}{Qingyun Wu}, \bibinfo{person}{Gagan Bansal}, \bibinfo{person}{Jieyu Zhang}, \bibinfo{person}{Yiran Wu}, \bibinfo{person}{Shaokun Zhang}, \bibinfo{person}{Erkang Zhu}, \bibinfo{person}{Beibin Li}, \bibinfo{person}{Li Jiang}, \bibinfo{person}{Xiaoyun Zhang}, {and} \bibinfo{person}{Chi Wang}.} \bibinfo{year}{2023}\natexlab{}.
\newblock \showarticletitle{Autogen: Enabling next-gen llm applications via multi-agent conversation framework}.
\newblock \bibinfo{journal}{\emph{arXiv preprint arXiv:2308.08155}} (\bibinfo{year}{2023}).
\newblock


\bibitem[Yuan et~al\mbox{.}(2024)]%
        {yuan2024easytool}
\bibfield{author}{\bibinfo{person}{Siyu Yuan}, \bibinfo{person}{Kaitao Song}, \bibinfo{person}{Jiangjie Chen}, \bibinfo{person}{Xu Tan}, \bibinfo{person}{Yongliang Shen}, \bibinfo{person}{Ren Kan}, \bibinfo{person}{Dongsheng Li}, {and} \bibinfo{person}{Deqing Yang}.} \bibinfo{year}{2024}\natexlab{}.
\newblock \showarticletitle{Easytool: Enhancing llm-based agents with concise tool instruction}.
\newblock \bibinfo{journal}{\emph{arXiv preprint arXiv:2401.06201}} (\bibinfo{year}{2024}).
\newblock


\bibitem[Zamfirescu and Hartmann(2023)]%
        {zamfirescu2023iterative}
\bibfield{author}{\bibinfo{person}{Raluca Zamfirescu} {and} \bibinfo{person}{Bjoern Hartmann}.} \bibinfo{year}{2023}\natexlab{}.
\newblock \showarticletitle{Iterative Disambiguation: Towards LLM-Supported Programming and System Design}. In \bibinfo{booktitle}{\emph{ICML Workshop on Interpretable Machine Learning}}.
\newblock
\urldef\tempurl%
\url{https://people.eecs.berkeley.edu/~bjoern/papers/zamfirescu-iterdis-icmlws2023.pdf}
\showURL{%
\tempurl}


\bibitem[Zhao et~al\mbox{.}(2023)]%
        {zhao2023survey}
\bibfield{author}{\bibinfo{person}{Wayne~Xin Zhao}, \bibinfo{person}{Kun Zhou}, \bibinfo{person}{Junyi Li}, \bibinfo{person}{Tianyi Tang}, \bibinfo{person}{Xiaolei Wang}, \bibinfo{person}{Yupeng Hou}, \bibinfo{person}{Yingqian Min}, \bibinfo{person}{Beichen Zhang}, \bibinfo{person}{Junjie Zhang}, \bibinfo{person}{Zican Dong}, {et~al\mbox{.}}} \bibinfo{year}{2023}\natexlab{}.
\newblock \showarticletitle{A survey of large language models}.
\newblock \bibinfo{journal}{\emph{arXiv preprint arXiv:2303.18223}} (\bibinfo{year}{2023}).
\newblock


\end{thebibliography}

\end{document}